\documentclass[journal]{IEEEtran}
\usepackage{graphicx}
\usepackage[font=footnotesize]{caption}
\usepackage{cite}
\usepackage{color,soul}
\usepackage[none]{hyphenat}
\usepackage{hyperref}
\usepackage[cmex10]{amsmath}
\usepackage{mdwmath}
\usepackage{amssymb}
\usepackage{dsfont}
\usepackage{subfig}
\usepackage{multirow}
\usepackage{epstopdf}
\usepackage{color}

\ifCLASSINFOpdf

\else

\fi

\begin{document}
	
	\title{Impact of Device Orientation on Error Performance of LiFi Systems}

	\author{\IEEEauthorblockN{Mohammad Dehghani Soltani,~\IEEEmembership{ Student Member,~IEEE,} Ardimas Andi Purwita, ~\IEEEmembership{Student Member,~IEEE,}\\
			Iman Tavakkolnia, ~\IEEEmembership{Member,~IEEE}, 
			Harald Haas,~\IEEEmembership{Fellow,~IEEE,}} and
		Majid Safari,~\IEEEmembership{Member,~IEEE}
		\thanks{
			The authors are with the LiFi Research and Development Centre, Institute for Digital Communications, The University of Edinburgh, UK. (e-mail: \{m.dehghani, a.purwita, i.tavakkolnia, h.haas, majid.safari\}@ed.ac.uk).}
		
		
		\vspace{-0.6cm}
	}


	\maketitle
	
	\begin{abstract}
		Most studies on optical wireless communications (OWCs) have neglected the effect of random orientation in their performance analysis due to the lack of a proper model for the random orientation. Our recent empirical-based research illustrates that the random orientation follows a Laplace distribution for a static user equipment (UE). In this paper, we analyze the device orientation and assess its importance on system performance. 
		The reliability of an OWC channel highly depends on the availability and alignment of line-of-sight (LOS) links.
		In this study, the effect of receiver orientation including both polar and azimuth angles on the LOS channel gain are analyzed.
		The probability of establishing a LOS link is investigated and the probability density function (PDF) of signal-to-noise ratio (SNR) for a randomly-oriented device is derived. By means of the PDF of SNR, the bit-error ratio (BER) of DC-biased optical orthogonal frequency division multiplexing (DCO-OFDM) in additive white Gaussian noise (AWGN) channels is evaluated. A closed-form approximation for the BER of UE with random orientation is presented which shows a good match with Monte-Carlo simulation results. Furthermore, the impact of the UE's random motion on the BER performance has been assessed. Finally, the effect of random orientation on the average signal-to-interference-plus-noise ratio (SINR) in a multiple access points (APs) scenario is investigated.  
	\end{abstract}
	\vspace{-0.1cm}

	\IEEEpeerreviewmaketitle
	
	\vspace{-0.25cm}
	
	\section{Introduction}
	Statistical data traffic confirms that smartphones will generate more than $86\%$ percent of the total mobile data traffic by $2021$ \cite{Cisco}. Light-Fidelity (LiFi) as part of the future fifth generation can cope with this immense volume of data traffic  \cite{Haas}. 
	LiFi is a bidirectional networked system that utilizes visible light spectrum in the downlink and infrared spectrum in the uplink \cite{MDSFeedback}. LiFi offers remarkable advantages such as utilizing a very large and unregulated bandwidth, energy efficiency and enhanced security. These benefits have put LiFi in the scope of recent and future research \cite{Iman2018Energy}. 
	The majority of studies on optical wireless communications assume that the device always faces vertically upwards. Although this may be for the purpose of analysis simplification or due to lack of a proper model for device orientation, in a real life scenario users hold their device in a way that feels most comfortable. Device orientation can affect the users' throughput remarkably and it should be analyzed carefully. 
	Even though a number of studies have considered the impact of random orientation in their analysis \cite{APselection,MDSHandover,ArdimasVTC2018,ImpactTilted,ICCTilting,wang2011performance,le2015impact,matrawy2016optimum,ErogluArxiv2017Orientation}. Device orientation can be measured by the gyroscope and accelerator implemented in every smartphone \cite{androidapp}. Then, this information can be fedback to the access point (AP) by the limited-feedback schemes to enhance the system throughput \cite{MDSFeedback,soltani2017throughput,dehghani2015limited}. 
	
	The effect of random orientation on users' throughput has been assessed in \cite{APselection}. In order to tackle the problem of load balancing, the authors proposed a novel AP selection algorithm that considers the random orientation of user equipments (UEs). The downlink handover problem due to the random rotation of UE in LiFi networks is characterized in \cite{MDSHandover}. The handover probability and handover rate for static and mobile users are determined. The handover probability in hybrid LiFi/RF-based networks with randomly-oriented UEs is analyzed in \cite{ArdimasVTC2018}. The effect of tilting the UE on the channel capacity is studied and the lower and upper bounds of the channel capacity are derived in \cite{ImpactTilted}. A theoretical expression of the bit-error ratio (BER) using on-off keying (OOK) has been derived in \cite{ICCTilting}. Then, a convex optimization problem is formulated based on the derived BER expression to minimize the BER performance by tilting the UE plane properly.
	A similar approach is used in \cite{wang2011performance} by finding the optimal tilting angle to improve both the signal-to-noise ratio (SNR) and spectral efficiency of M-QAM orthogonal frequency division multiplexing (OFDM) for indoor visible light communication (VLC) systems. 
	Impacts of both UE's orientation and position on link performance of VLC are studied in \cite{le2015impact}. The outage probability is derived and the significance of UE orientation on inter-symbol interference is shown. 
	The optimum polar and azimuth angles for single user multiple-input multiple-output (MIMO) OFDM is calculated in \cite{matrawy2016optimum}. A receiver with four photodetectors (PD) is considered and the optimal angles for each PD are computed. 
	In \cite{ErogluArxiv2017Orientation}, the impact of the random orientation on the line-of-sight (LOS) channel gain for a randomly located UE is studied. The statistical distribution of the channel gain is presented for a single light-emitting diode (LED) and extended to a scenario with double LEDs.  
	All mentioned studies assume a predefined model for the random orientation of the receiver. However, little or no evidence is presented to justify the assumed models. For the first time, experimental measurements are carried out to model the polar and azimuth angles in \cite{MDSArxiv2018Orientation,ArdimasWCNC2018,ZhihongOrientation}. It is shown that the polar angle can be modeled by either the Laplace distribution (for static users) or the Gaussian distribution (for mobile users) while the azimuth angle follows a uniform distribution. Solutions to alleviate the impact of device random orientation on received SNR and throughput are proposed in \cite{MDSJSAC18,ImanICCW19,ChengICCW19}. 
	In \cite{MDSJSAC18}, the statistics of Euler rotation angles are provided based on the experimental measurements. Then, simulations of BER performance for spatial modulation using a multi-directional receiver configuration with consideration of random device orientation is evaluated. In \cite{ImanICCW19}, other multiple-input multiple-output (MIMO) techniques in the presence of random orientation are studied. 
	The authors in \cite{ChengICCW19}, proposed an omni-directional receiver which is not affected by device random orientation. It is shown that the omni-directional receiver reduces the SNR fluctuations and improves the user throughput remarkably. All these studies emphasize the significance of incorporating the random orientation into the analysis.
	
	We characterize the device random orientation and investigate its effect on the users' performance metrics such as SNR and BER in optical wireless systems. We also derive the probability density function (PDF) of SNR for randomly-orientated device. 
	Based on the derived PDF of SNR, the BER performance of a DC biased optical OFDM (DCO-OFDM) is evaluated as a use case. A closed form approximation for BER is purposed. The impact of device orientation on BER with some interesting observations are investigated. In this study, we only consider the LOS channel gain, and the impact of higher reflections on BER performance has been investigated in our recent study \cite{ArdimasOFDM}. 
	\vspace{-0.0cm}
	

	\textit{Notations:} $|\cdot|$ expresses the absolute value of a variable;  $\tan^{-1}(y/x)$ is the four-quadrant inverse tangent. Further, $[\cdot]^{\rm{T}}$ stands for transpose operator. We note that throughout this paper, unless otherwise mentioned, angles are expressed in degrees. The Gaussian distribution with mean, $\mu_{\rm G}$, and variance, $\sigma^2_{\rm G}$, is denoted by $\mathcal{N}(\mu_{\rm G},\sigma^2_{\rm G})$.
	
	\begin{figure}[t!]
		\centering
		\resizebox{0.9\linewidth}{!}{
			\includegraphics{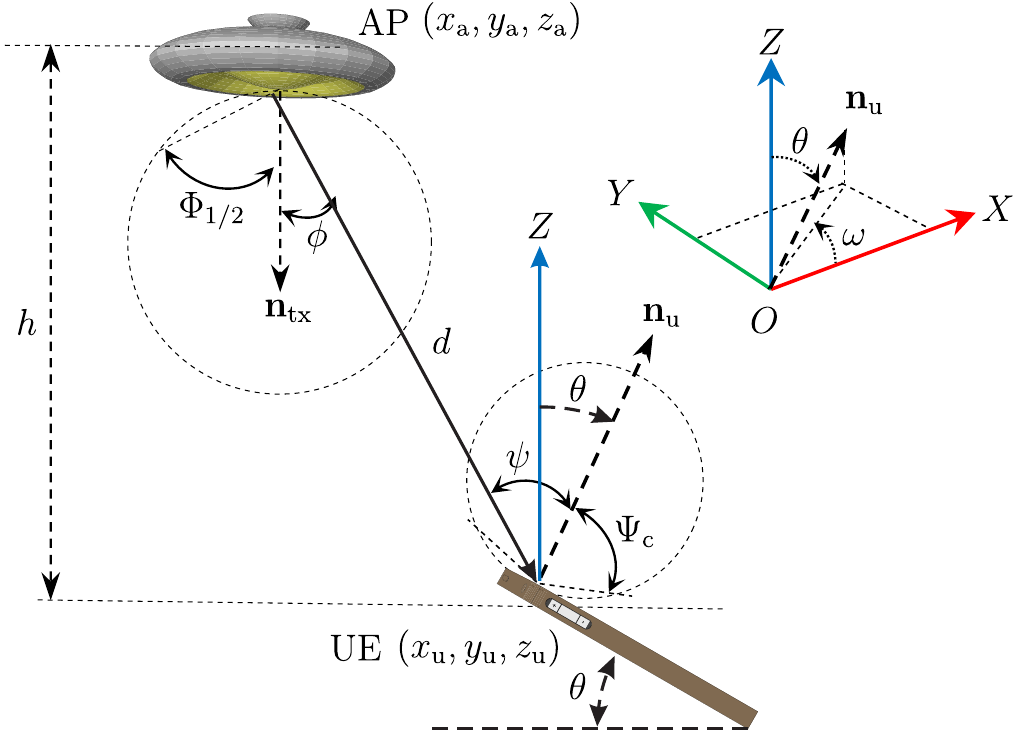}}
		\caption{Downlink geometry of light propagation in LiFi networks.}  
		\label{FigSystemModel} 
		\vspace{-0.3cm}
	\end{figure}
	
	\section{System Model}
	\label{SystemModel}
	\subsection{LOS Channel Gain}
	An open indoor office without reflective objects for optical wireless downlink transmission is considered in this study. The geometric configuration of the downlink transmission is illustrated in Fig.~\ref{FigSystemModel}. It is assumed that an LED transmitter (or AP) is a point source that follows the Lambertian radiation pattern. Furthermore, the LED is supposed to operate within the linear dynamic range of the current-power characteristic curve to avoid the nonlinear distortion effect. The LED is fixed and oriented vertically downward. 

	The direct current (DC) gain of the LOS optical wireless channel between the AP and the UE is given by \cite{Kahn}:
	\begin{equation}
	\label{LOSChnlGain}
	H=\frac{(m+1)A_{\rm{PD}}}{2\pi d^2}g_{\rm{f}}\cos^m\phi\cos\psi\ {\rm{rect}\left( \frac{\psi}{\Psi_{\rm c}}\right) },
	\end{equation}
	where ${\rm{rect}(\frac{\psi}{\Psi_{\rm c}})}=1$ for $0\leq\! \psi \!\leq\! \Psi_{\rm c}$ and $0$ otherwise; $A_{\rm{PD}}$ is the PD physical area; the Euclidean distance between the AP and the UE is denoted by $d$ with $(x_{\rm{a}},y_{\rm{a}},z_{\rm{a}})$ and $(x_{\rm{u}},y_{\rm{u}},z_{\rm{u}})$ as the position of the AP and UE in the Cartesian coordinate system, respectively; the Lambertian order is $m=-1/\log_2(\cos\Phi_{1/2})$ where $\Phi_{1/2}$ is the transmitter semiangle at half power. The incidence angle with respect to the normal vector to the UE surface, $\mathbf{n}_{\rm{u}}$, and the radiance angle with respect to the normal vector to the AP surface, $\mathbf{n}_{\rm{tx}}=[0, 0, -1]$, are denoted by $\phi$ and $\psi$, respectively. These two angles can be obtained by using the analytical geometry rules as $\cos\phi={\bf{d}}\cdot{\bf{n}}_{\rm{tx}}/d$ and $\cos\psi=-{\bf{d}}\cdot{\bf{n}}_{\rm{u}}/d$ where ${\bf{d}}$ is the distance vector from the AP to the UE and $``\cdot"$ is the inner product operator. The gain of the optical concentrator is given as $g_{\rm{f}}=\varsigma^2 /\sin^2\Psi_{\rm{c}}$ with $\varsigma$ being the refractive index and $\Psi_{\rm{c}}$ is the UE field of view (FOV). After some simplifications, \eqref{LOSChnlGain} can be written as:
	\begin{equation}
	\label{LOSSimple}
	H=\frac{H_0\cos\psi}{d^{m+2}}{\rm{rect}\left( \frac{\psi}{\Psi_{\rm c}}\right) },
	\end{equation}
	where $H_0=\frac{(m+1)A_{\rm{PD}}g_{\rm{f}}h^{m}}{2\pi}$; and $h=|z_{\rm{a}}-z_{\rm{u}}|$ is the vertical distance between the UE and the AP as shown in Fig.~\ref{FigSystemModel}.
	\vspace{-0.1cm}

	\begin{figure}[t!]
		\centering
		\subfloat[Normal position\label{sub1:fig2}]{%
			\includegraphics[width=40mm,height=35mm]{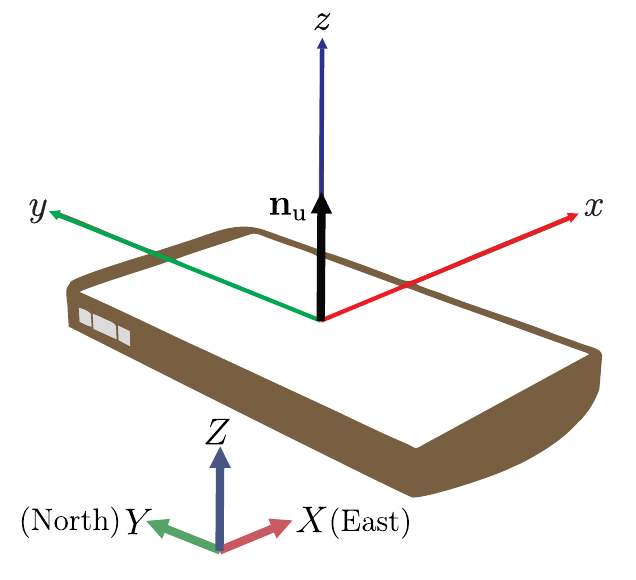}
		}
		\quad\vspace{-.3cm}
		\subfloat[Yaw rotation with angle $\alpha$\label{sub2:fig2}]{%
			\includegraphics[width=40mm,height=35mm]{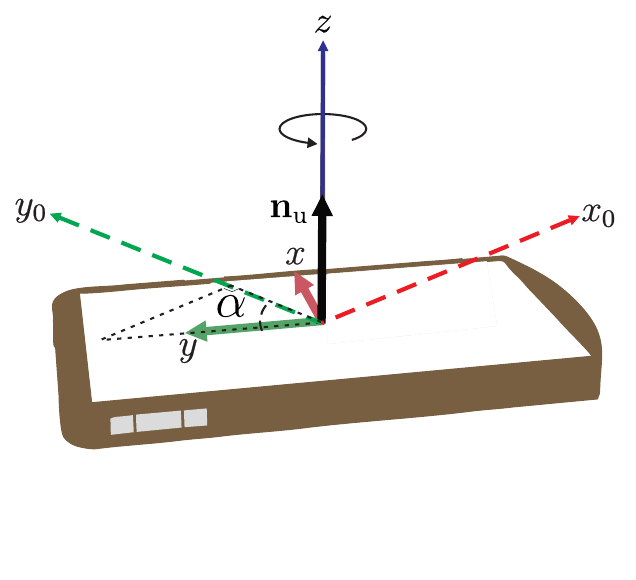}
		}\ 
		\subfloat[Pitch rotation with angle $\beta$\label{sub3:fig2}]{%
			\includegraphics[width=40mm,height=35mm]{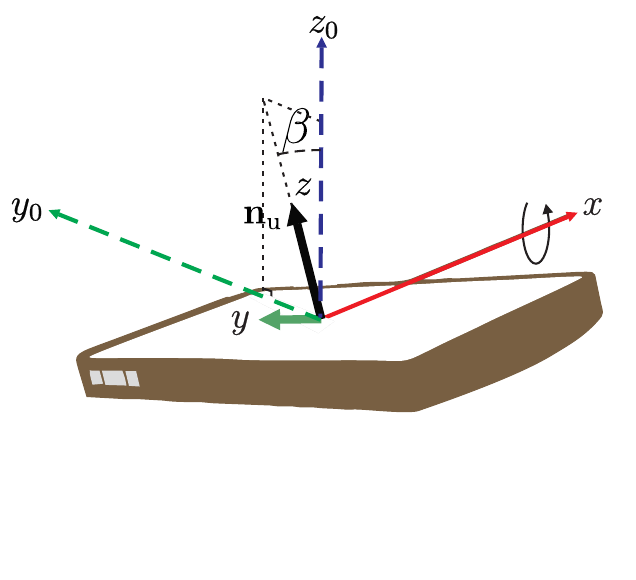}
		}
		\quad
		\subfloat[Roll rotation with angle $\gamma$\label{sub4:fig2}]{%
			\includegraphics[width=40mm,height=35mm]{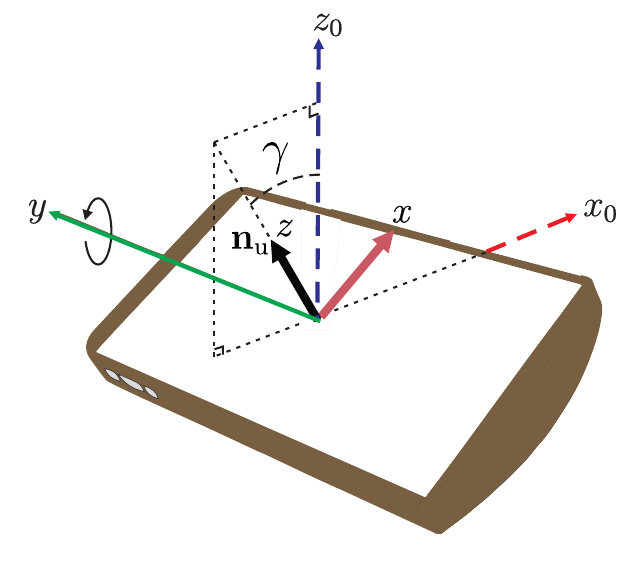}
		}\\ \vspace{-1pt}
		\caption{Orientations of a mobile device \cite{MDSArxiv2018Orientation}.}
		\label{FigOrientation}
		\vspace{-8pt}
	\end{figure}

	\addtocounter{equation}{1}
	\begin{figure*}[th!]
		\begin{align} 
		\label{RotatedNormalVec}
		\mathbf{n}'_{\rm{u}}\!=\! 
		\mathbf{R}_\alpha \mathbf{R}_\beta \mathbf{R}_\gamma \!
		\begin{bmatrix}
		0 \\
		0 \\
		1 
		\end{bmatrix}\!\!\!
		&=\!\!
		\begin{bmatrix}
		\cos \alpha\! & -\sin \alpha\! & 0\! \\
		\sin \alpha\! & \cos \alpha\! & 0\! \\
		0\! & 0\! & 1\! 
		\end{bmatrix}\!\!\!
		\begin{bmatrix}
		1\! & 0\! & 0\! \\
		0\! & \cos \beta\! & -\sin \beta\! \\
		0\! & \sin \beta\! & \cos \beta \!
		\end{bmatrix}\!\!\!
		\begin{bmatrix}
		\cos \gamma\! & 0\! & \sin \gamma\! \\
		0\! & 1\! & 0\! \\
		-\sin \gamma\! & 0\! & \cos \gamma\! 
		\end{bmatrix} \!\!\!
		\begin{bmatrix}
		0 \\
		0 \\
		1 
		\end{bmatrix}
		\!\!=\!\! 
		\begin{bmatrix}
		\cos \gamma \sin \alpha \sin \beta+\cos \alpha \sin \gamma \\
		\sin \alpha \sin \gamma-\cos \alpha \cos \gamma \sin \beta \\
		\cos \beta \cos \gamma 
		\end{bmatrix}.
		\end{align} 
		\vspace{-0.1cm}
		\hrule
		\vspace{-0.2cm}
	\end{figure*}
	
	\subsection{Rotation in the Space}
	\label{Orientation}
	
	A convenient way of describing the orientation is to use three separate angles showing the rotation about each axes of the rotating local coordinate system (intrinsic rotation) or the rotation about the axes of the reference coordinate system (extrinsic rotation). 
	Current smartphones are able to report the elemental intrinsic rotation angles yaw, pitch and roll denoted as $\alpha$, $\beta$ and $\gamma$, respectively \cite{Barthold}. Here, $\alpha$ represents rotation about the $z$-axis, which takes a value in range of $[0,360)$; $\beta$ denotes the rotation angle about the $x$-axis, that is, tipping the device toward or away from the user, which takes value between $-180^\circ$ and $−180^\circ$; and $\gamma$ is the rotation angle about the $y$-axis, that is, tilting the device right or left, which is chosen from the range $[-90,90)$. The elemental Euler angles are depicted in Fig.~\ref{FigOrientation}.

	
	Now we derive the concatenated rotation matrix with respect to the reference coordinate system. The normal vector after rotation can be obtained as:\vspace{-0.1cm}
	\newcounter{MYtempeqncnt}
	\begin{equation}
	\setcounter{MYtempeqncnt}{\value{equation}}
	\setcounter{equation}{3}
	\label{NormalVecRot}
	\mathbf{n}'_{\rm{u}}=\mathbf{R}\mathbf{n}_{\rm{u}},
	\vspace{-0.0cm}
	\end{equation} 
	where $\mathbf{n}_{\rm{u}}=[n_1, n_2, n_3]^{\rm{T}}$ is the original normal vector and $\mathbf{n}'_{\rm{u}}=[n'_1, n'_2, n'_3]^{\rm{T}}$ is the rotated normal vector via the rotation matrix $\mathbf{R}$. The rotation matrix can be decomposed as $\mathbf{R}=\mathbf{R}_\alpha\mathbf{R}_\beta\mathbf{R}_\gamma$, where $\mathbf{R}_\alpha$, $\mathbf{R}_\beta$ and $\mathbf{R}_\gamma$ are the rotation matrices about the $z$, $x$ and $y$ axes, respectively. Assume that the body frame and the reference frame are initially aligned so that $\mathbf{n}_{\rm{u}}=[0, 0, 1]^{\rm{T}}$, then, the rotated normal vector, $\mathbf{n}'_{\rm{u}}$, via the rotation matrices $\mathbf{R}_\alpha$, $\mathbf{R}_\beta$ and $\mathbf{R}_\gamma$ is given in \eqref{RotatedNormalVec} shown at top this page. 
	
	\begin{figure}[t!]
		\begin{center}
			\resizebox{0.8\linewidth}{!}{\includegraphics{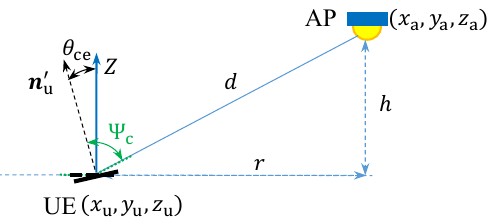}}
			\caption{Geometry of critical elevation angle.}
			\vspace{-0.3cm}
			\label{Angles}
		\end{center}
	\end{figure}
	
	The rotated normal vector can be represented in the spherical coordinate system using the azimuth, $\omega$, and polar, $\theta$ angles. That is, $\mathbf{n}'_{\rm{u}}=[\sin\theta\cos\omega,\sin\theta\sin\omega,\cos\theta]^{\rm{T}}$. As shown in Fig.~\ref{FigSystemModel}, $\theta$ is the angle between the positive direction of the $Z$-axis and the normal vector $\mathbf{n}'_{\rm{u}}$, also $\omega$ is the angle between the projection of $\mathbf{n}'_{\rm{u}}$ in the $XY$-plane and the positive direction of the $X$-axis. Accordingly, 
	\begin{equation}
	\setcounter{equation}{5}
	\begin{aligned}
	\label{ThetaAndOmega}
	&\theta = \cos^{-1} \left( \cos\beta \cos\gamma \right), \\
	&\omega\!=\!\tan^{-1}\!\left(\frac{n'_2}{n'_1} \right)\!\!=\! \tan^{-1}\!\left(\!\frac{\sin \alpha \sin \gamma-\cos \alpha \cos \gamma \sin \beta}{\cos \gamma \sin \alpha \sin \beta+\cos \alpha \sin \gamma} \right). 
	\end{aligned}
	\end{equation}
	It is shown in \cite{MDSArxiv2018Orientation} and \cite{ArdimasWCNC2018} that the elevation angle follows a Laplace distribution, $\theta\sim\mathcal{L}(\mu_{\theta},b_{\theta})$ where the mean value, $\mu_{\theta}$, and scale parameter, $b_{\theta}$, depend on whether the user is static or mobile. The mean is reported to be about $41^{\circ}$ and $30^{\circ}$ for sitting and walking activities, respectively \cite{MDSArxiv2018Orientation}. Furthermore, it is shown that the azimuth angle follows a uniform distribution, $\omega\sim\mathcal{U}[0,2\pi]$. For the rest of the paper, we consider the user's facing direction angle as $\Omega=\omega+\pi$, where $\Omega$ provides a better physical concept (compared to $\omega$), as it shows the angle between the user's facing direction and the $X$-axis.

	\section{Orientation Analysis}
	\label{Orientation Analysis}
	Before analyzing user's performance metrics such as average SNR and BER, let us define the critical elevation (CE), $\theta_{\rm{ce}}$, which defines the elevation angle at the boundary of the field of view of the receiver. As shown in Fig.~\ref{Angles}, the CE angle for a given position of UE, $(x_{\rm{u}},y_{\rm{u}})$, and user's direction, $\Omega$, is the elevation angle for which $\psi=\Psi_{\rm{c}}$. Thus, $\theta \geq \theta_{\rm{ce}}$ results in $\psi\geq\Psi_{\rm{c}}$, and the channel gain would be zero based on \eqref{LOSChnlGain}. This angle depends on both the UE position and its direction, $\Omega$ which is given as follows:\vspace{-0.1cm}
	\begin{equation}
	\label{CEComplete}
	\theta_{\rm{ce}}=\cos^{-1}\left(\frac{\cos\Psi_{\rm{c}}}{\sqrt{\lambda_1^2+\lambda_2^2}} \right)+\tan^{-1}\left(\frac{\lambda_1}{\lambda_2} \right),
	\end{equation}
	where the coefficients $\lambda_1$ and $\lambda_2$ are given as:
	\begin{equation}
	\begin{aligned}
	\label{Lambdas}
	&\lambda_1=\frac{r}{d}\cos\left( \Omega-\tan^{-1}\left(\frac{y_{\rm{u}}-y_{\rm{a}}}{x_{\rm{u}}-x_{\rm{a}}} \right) \right),\\
	&\lambda_2=\frac{h}{d}.
	\end{aligned}
	\end{equation}
	where $r=\sqrt{(x_{\rm{u}}-x_{\rm{a}})^2+(y_{\rm{u}}-y_{\rm{a}})^2}$ is the horizontal distance between the AP and the UE. 
	Proof of \eqref{CEComplete} is provided in Appendix-\ref{App-CE}. As can be seen from \eqref{Lambdas}, the parameter $\lambda_1$ contains the direction angle, $\Omega$. The physical concept of positive $\lambda_1$ is that the UE is facing to the AP while if it is not facing to the AP, $\lambda_1$ is negative. On the other hand, since always $z_{\rm{u}}<z_{\rm{a}}$, we have $\lambda_2>0$. It should be mentioned that the acceptable range for $\theta_{\rm{ce}}$ is $[0,90]$ as the polar angle, $\theta$, given in \eqref{ThetaAndOmega} takes values between $0^{\circ}$ and $90^{\circ}$. Note that for a given location of UE, the minimum CE angle, $\theta_{\rm th}$, is obtained for $\Omega=\pi+\tan^{-1}\left(\frac{y_{\rm{u}}-y_{\rm{a}}}{x_{\rm{u}}-x_{\rm{a}}} \right)\triangleq\Omega_{\rm{th}}$ which is given as:
	\begin{equation}
	\label{CE}
	\theta_{\rm th}=\Psi_{\rm{c}}+\sin^{-1}\left(\frac{h}{d} \right)-\frac{\pi}{2}. 
	\end{equation}
	
	

	\begin{figure}
		\begin{center}
			\resizebox{.9\linewidth}{!}{\includegraphics{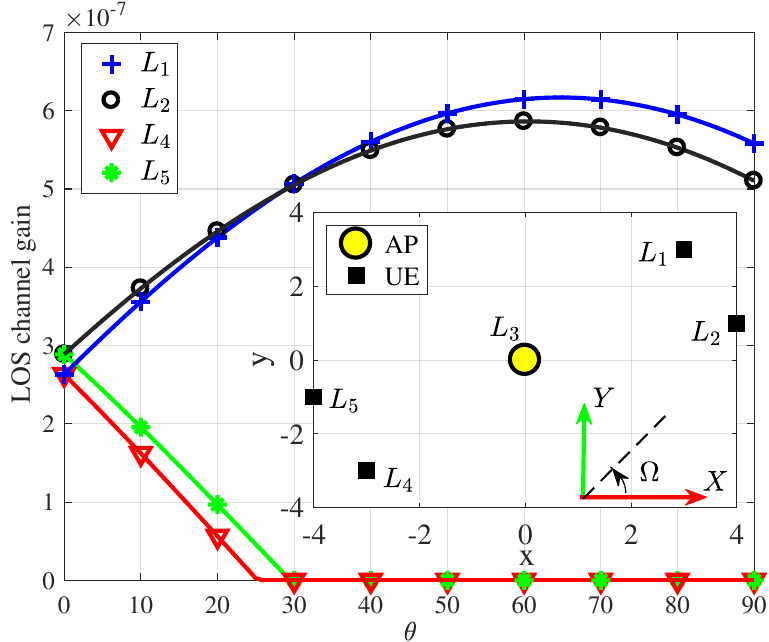}}
			\caption{The effect of changing $\theta$ on $\cos\psi$ for different locations of the UE with fixed $\Omega=45^\circ$ and $\Psi_{\rm{c}}=90^\circ$. }
			\vspace{-0.1cm}
			\label{figCEOmega}
		\end{center}
	\end{figure}

	\begin{table}[t]
		\centering
		\caption{Simulation Parameters}
		\label{TableSimulationParam}
		\vspace{-8pt}
		{\raggedright
			\vspace{4pt} \noindent
			\begin{tabular}{p{115pt}|p{40pt}|p{50pt}}
				\hline
				\parbox{115pt}{\centering{\small Parameter}} & \parbox{40pt}{\centering{\small Symbol}} & \parbox{50pt}{\centering{\small Value}} \\
				\hline
				\hline
				\parbox{115pt}{\raggedright{\small AP location }} & \parbox{40pt}{\centering{\small $(x_{\rm{a}},y_{\rm{a}},z_{\rm{a}})$}} & \parbox{50pt}{\centering{\small $(0,0,2)$}} \\
				\hline
				\parbox{115pt}{\raggedright{\small LED half-intensity angle}} & \parbox{40pt}{\centering{\small $\Phi_{1/2}$}} & \parbox{50pt}{\centering{\small $60^\circ$}} \\
				\hline
				\parbox{115pt}{\raggedright{\small PD responsivity}} & \parbox{40pt}{\centering{\small $R_{\rm{PD}}$}} & \parbox{50pt}{\centering{\small $1$ A/W }} \\
				\hline
				\parbox{115pt}{\raggedright{\small Physical area of a PD}} & \parbox{40pt}{\centering{\small $A_{\rm{PD}}$}} & \parbox{50pt}{\centering{\small $1$ cm$^2$}} \\
				\hline
				\parbox{115pt}{\raggedright{\small Refractive index}} & \parbox{40pt}{\centering{\small $\varsigma$}} & \parbox{50pt}{\centering{\small $1$}} \\
				\hline
				\parbox{115pt}{\raggedright{\small Downlink bandwidth}} & \parbox{40pt}{\centering{\small $B$}} & \parbox{50pt}{\centering{\small $10$ MHz}} \\
				\hline
				\parbox{115pt}{\raggedright{\small Number of subcarriers}} & \parbox{40pt}{\centering{\small $\mathcal{K}$}} & \parbox{50pt}{\centering{\small $1024$}} \\
				\hline
				\parbox{115pt}{\raggedright{\small Noise power spectral density}} & \parbox{40pt}{\centering{\small $N_0$}} & \parbox{50pt}{\centering{\small $10^{-21}$ A$^2$/Hz }} \\
				\hline
				\parbox{115pt}{\raggedright{\small Conversion factor}} & \parbox{40pt}{\centering{\small $\eta$}} & \parbox{50pt}{\centering{\small $3$ }} \\
				\hline
				\parbox{120pt}{\raggedright{\small Vertical distance of UE and AP}} & \parbox{40pt}{\centering{\small $h$}} & \parbox{50pt}{\centering{\small $2$ m}} \\
				\hline
			\end{tabular}
			\vspace{2pt}
		}
		\vspace{-8pt}
	\end{table}
	

	\begin{figure}[t!]
		\centering  
		\includegraphics[width=0.95\columnwidth,height=65mm,trim={0.5cm 0 -0.5cm 0}]{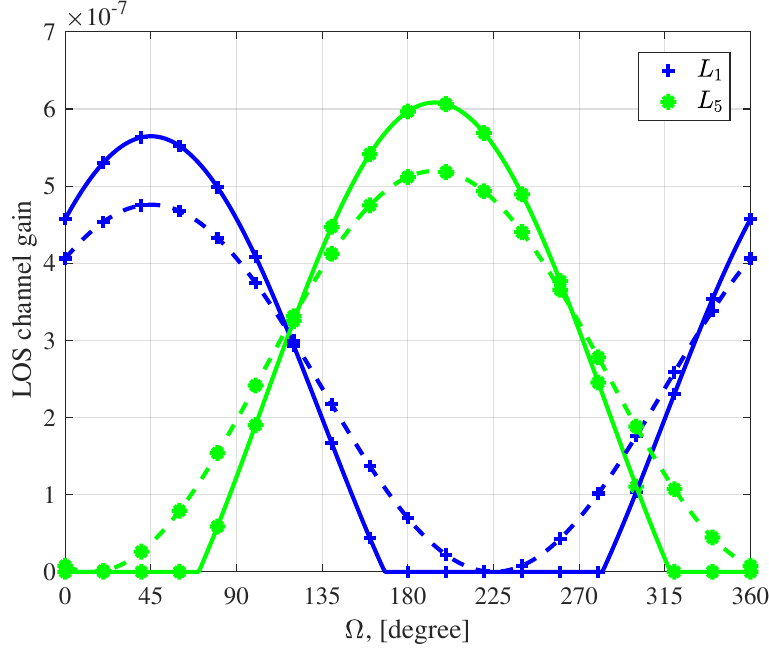} 
		\caption{The effect of changing $\Omega$ and $\theta$ on the LOS channel gain with $\Psi_{\rm{c}}=90^{\circ}$, for different positions and elevation angles $\theta=41^{\circ}$ (solid lines), $\theta=\theta_{\rm th}$ (dash lines). }
		\label{figCEEffect}
		\vspace{-0.2cm}
	\end{figure}

	The effect of changing the elevation angle, $\theta$, on the LOS channel gain for different locations of the UE with a fixed direction angle, $\Omega\!=\!45^\circ$ and $\Psi_{\rm{c}}=90^\circ$, is shown in Fig.~\ref{figCEOmega}. Here, $\Psi_{\rm{c}}\!=\!90^\circ$ and other parameters are presented in Table~\ref{TableSimulationParam}. It can be seen that for the UE's locations of $L_4\!=\!(-3,-3)$ and $L_5\!=\!(-4,-1)$ by increasing the elevation angle, the LOS channel gain decreases. After $\theta_{\rm{ce}}=25.24^\circ$ and $\theta_{\rm{ce}}=29.5^\circ$ for $L_4$ and $L_5$, respectively, the AP is out of the UE's FOV and hence the LOS channel gains are zero. However, with the same $\Omega\!=\!45^\circ$ if the UE is located at positions like $L_1\!=\!(3,3)$ or $L_2\!=\!(4,1)$, the LOS channel gain does not become zero if the elevation angle changes between $0^\circ$ and $90^\circ$. 


	It is noted that under the condition of $\theta<\theta_{\rm{th}}$ the AP is always within the UE's field of view for any direction of $\Omega$. 
	For a given UE's location, we are also interested in the range of $\Omega$ for which the LOS channel is active. Let's denote this range as $\mathcal{R}_{\Omega,\theta}$. This range can be determined according to the following Proposition.

	\textit{\textbf{Proposition.}} For a given UE's location, the range of $\Omega$ for which the LOS channel gain is non-zero is $[0, 2\pi]$ if $\theta$ is smaller than or equal to a threshold angle $\theta_{\rm th}=\Psi_{\rm{c}}+\sin^{-1}\left(\frac{h}{d} \right)-\frac{\pi}{2}$. Otherwise it is given as follows:\vspace{-0.0cm}
	\begin{equation}
	\mathcal{R}_{\Omega,\theta}\!=\!\begin{cases}
	[0,\Omega_{\rm{r1}})\bigcup(\Omega_{\rm{r2}},2\pi],  \ \rm{if}\ \Lambda'(\Omega_{\rm{r1}})<0 \\
	(\Omega_{\rm{r1}},\Omega_{\rm{r2}}), \ \ \ \ \ \ \ \ \ \ \ \ \ \rm{if}\ \Lambda'(\Omega_{\rm{r1}})\geq 0
	\end{cases},
	\end{equation}
	where $\Lambda'(\Omega)=-\kappa_1\sin\left(\Omega-\tan^{-1}\left(\frac{y_{\rm{u}}-y_{\rm{a}}}{x_{\rm{u}}-x_{\rm{a}}} \right)\right)+\kappa_2$ with:\vspace{-0.1cm}
	\begin{equation}
	\label{kappas}
	\begin{aligned}
	&\kappa_1=\frac{r}{d}\sin\theta, \ \ \ \ \ 
	&\kappa_2=\frac{h}{d}\cos\theta \ .
	\end{aligned}
	\end{equation}
	Also $\Omega_{\rm{r1}}=\min\{\Omega_1,\Omega_2\}$ and $\Omega_{\rm{r2}}=\max\{\Omega_1,\Omega_2\}$, where:
	\begin{equation}
	\label{RootsPr}
	\begin{aligned}
	&\Omega_1=\cos^{-1}\!\left(\!\frac{\cos\Psi_{\rm{c}}-\kappa_2}{\kappa_1}\!\right)+\!\tan^{-1}\!\!\left(\!\frac{y_{\rm{u}}-y_{\rm{a}}}{x_{\rm{u}}-x_{\rm{a}}}\! \right),\\
	&\Omega_2=-\cos^{-1}\!\left(\!\frac{\cos\Psi_{\rm{c}}-\kappa_2}{\kappa_1}\!\right)+\!\tan^{-1}\!\!\left(\!\frac{y_{\rm{u}}-y_{\rm{a}}}{x_{\rm{u}}-x_{\rm{a}}}\! \right).
	\end{aligned}
	\end{equation}
	
	\textit{Proof: See Appendix-\ref{App-Proposition}}. 
	
	\begin{figure}[t!]
		\centering  
		\includegraphics[width=0.95\columnwidth,height=65mm,trim={0.5cm 0 -0.5cm 0}]{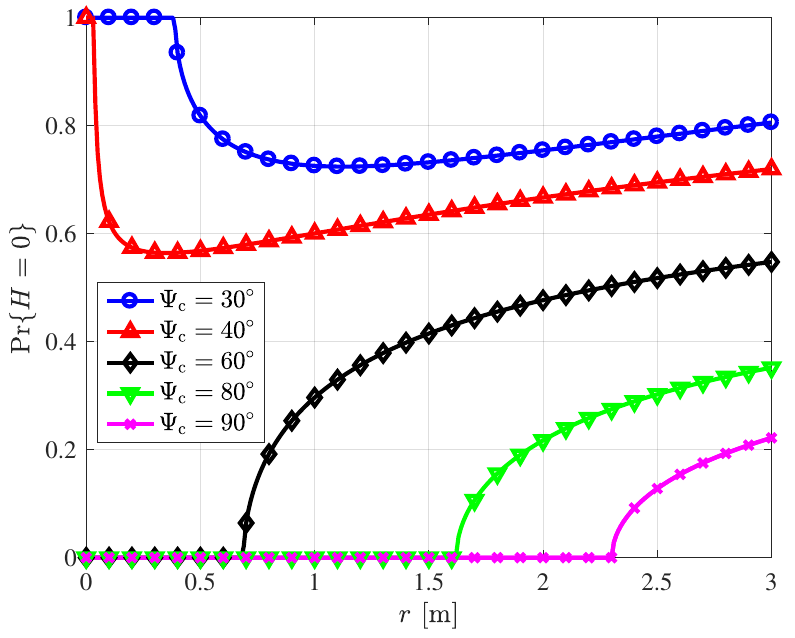} 
		\caption{The effect of different FOV on having a zero LOS, $\Pr\{H=0\}$. }
		\label{figLOSoutage}
		\vspace{-0.2cm}
	\end{figure}
	

	The LOS channel gain versus $\Omega$ for locations of $L_1$ and $L_5$ (see the inset of Fig.~\ref{figCEOmega}) with $\theta=\theta_{\rm th}$ (dash line) and $\theta=41^{\circ}\geq\theta_{\rm th}$ (solid line) are shown in Fig.~\ref{figCEEffect}. Note that for $L_1$ and $L_5$, we have $\theta_{\rm th}=25.24^{\circ}$ and $\theta_{\rm th}=25.88^{\circ}$, respectively. As can be seen, if $\theta<\theta_{\rm th}$, then, $\forall \Omega\in[0,360)$, LOS channel gain is always non-zero (dash lines). Based on the proposition, the range of $\Omega$ for which the LOS channel gain is non-zero with $\theta=41^\circ>\theta_{\rm th}$ is $[0,167.8)\cup(282.2,360]$ for $L_1$ and $[70.1,318]$ for $L_2$. 
	
	It can be inferred form the Proposition that for a given UE's location and $\theta$, the probability that the LOS path is not within the UE's FOV (due to variation of $\Omega$) is $\Pr\{H=0\}=1-\Pr\{\Omega\in\mathcal{R}_{\Omega,\theta}\}$. Fig.~\ref{figLOSoutage} shows the $\Pr\{H=0\}$ versus the horizontal distance between the UE and the AP, $r$, for different UE's FOV. The results are shown for $\theta=41^{\circ}$. As can be observed, $\Pr\{H=0\}=1$, for UEs with a narrow FOV (i.e., $\Psi_{\rm c}=30^{\circ}$ and $40^{\circ}$) when they are located in the vicinity below the AP. As the horizontal distance, $r$, increases, $\Pr\{H=0\}$ first decreases and then it increases as it goes away from the AP. For wide FOVs (i.e., $\Psi_{\rm c}=60^{\circ}$, $80^{\circ}$ and $90^{\circ}$), $\Pr\{H=0\}$ is zero when the UE is in the vicinity below the AP, and then it starts to increase at a certain $r$. This can be derived based on \eqref{CE} for $\Psi_{\rm c}\geq \theta$ that is $r\geq h\tan(\Psi_{\rm c}-\theta)$. Note that the high value of losing the LOS link particularly for narrower FOVs is due the fact that a single AP is considered and the effect of reflection is ignored. A study of such effects has been presented in our recent work \cite{ArdimasOFDM}.   
	
	Let $\mathcal{R}_{\Omega}$ denote the range for which the LOS channel gain is always non-zero regardless of $\theta$, i.e., $\forall \theta\in[0,90]$. The range, $\mathcal{R}_{\Omega}$, can be determined according to the following Corollary.

	\textit{\textbf{Corollary.}}
	For a given UE's location, the range of $\Omega$ for which the LOS channel gain is non-zero for all $\theta\in [0,90]$ can be obtained as:
	\begin{equation}
	\label{Rtheta}
	\mathcal{R}_{\Omega}=\mathcal{R}_{\Omega,\theta}|_{\rm{For}\ \theta=90}. 
	\end{equation}
	Proof of this corollary is similar to the proof of proposition 1. Noting that the worst elevation angle that leads to the smallest range of $\Omega$ is $\theta=90^\circ$. 
	The physical concept of $\mathcal{R}_{\Omega}$ is that when the UE faces the AP, we have $\Omega\in\mathcal{R}_{\Omega}$. Otherwise, if the UE faces the opposite direction of the AP, $\Omega\notin\mathcal{R}_{\Omega}$.
	In fact, $\mathcal{R}_{\Omega}$ provides a stable range for which the user can change the elevation angle between $0$ and $90$ without experiencing the AP out of its FOV.  
	We note that the range given in \eqref{Rtheta} is valid if $\Psi_{\rm{c}}\geq\cos^{-1}\left(\frac{r}{d}\right)$ (this condition can be readily seen by substituting $\theta=90^{\circ}$ in \eqref{kappas} and then replacing the results in \eqref{RootsPr}).

	\section{Bit-Error Ratio Performance}
	\label{BERPerf}
	In this section, we evaluate the BER performance of DCO-OFDM in LiFi networks. We initially derive the SNR statistics on each subcarrier, then based on the derived PDF of SNR, the BER performance is assessed. 
	Note that the PDF of the SNR derived in this study is the conditional PDF given the location and direction of the UE. Therefore, having the statistics of the user location, the joint PDF of the SNR with respect to both UE orientation and location can be readily obtained.
	\subsection{SNR Statistics}
	The received electrical SNR\footnote{Note that all SNR values throughout this paper are scalers, i.e., not in dB.} on $k$th subcarrier of a LiFi system can be acquired as:
	\begin{equation}
	\label{SNR}
	\mathcal{S}=\frac{R_{\rm{PD}}^2H^2P_{\rm{opt}}^2}{(\mathcal{K}-2)\eta^2\sigma_k^2},
	\end{equation}
	where the PD responsivity is denoted by $R_{\rm{PD}}$; $H$ is the LOS channel gain given in \eqref{LOSChnlGain}; $P_{\rm{opt}}$ is the transmitted optical power; $\mathcal{K}$ is the total number of subcarriers with $\mathcal{K}/2-1$ subcarriers bearing information. Furthermore, $\eta$ is the conversion factor \cite{dissanayake2013comparison}. The condition $\eta=3$ can guarantee that less than $1\%$ of the signal is clipped so that the clipping noise is negligible \cite{DimitrovVTC,MDSFeedback}. In \eqref{SNR}, $\sigma_k^2=N_0B/\mathcal{K}$ is the noise power on $k$th subcarrier where $N_0$ stands for the noise spectral density and $B$ represents the modulation bandwidth. Based on the experimental measurement of the device orientation, it is shown in \cite{MDSArxiv2018Orientation} that the LOS channel gain, $H$, follows a clipped Laplace distribution as:
	\begin{equation}
	\label{LOSPDF}
	\begin{aligned}
	f_{\rm{H}}(\hbar)\!
	=\!\frac{\exp\left(-\frac{|\hbar-\mu_{\rm{H}}|}{b_{\rm{H}}} \right) }{b_{\rm{H}}\left(\!2-\exp\left(-\frac{h_{\rm{max}}-\mu_{\rm{H}}}{b_{\rm{H}}} \right)\!  \right) }\!+c_{\rm{H}}\delta(\hbar), 
	\end{aligned}
	\end{equation}
	where $\delta(\hbar)$ is the Dirac delta function, taking $1$ if $\hbar=0$, and $0$ otherwise; $c_{\rm{H}}=F_{\rm{\cos\psi}}(\cos\Psi_{\rm{c}})$, which is given as:
	\begin{equation}
	\label{delta_val}
	c_{\rm{H}}=F_{\rm{\cos\psi}}(\cos\Psi_{\rm{c}})\approx \begin{cases}
	1-\frac{1}{2}\exp\left( \frac{\theta_{\rm{ce}}-\mu_{\theta}}{b_{\theta}}\right),\ \   &\theta_{\rm{ce}}<\mu_{\theta}\\
	\frac{1}{2}\exp\left( -\frac{\theta_{\rm{ce}}-\mu_{\theta}}{b_{\theta}}\right), \ \ &\theta_{\rm{ce}}\geq\mu_{\theta} 
	\end{cases} \ .
	\end{equation}
	where $b_{\rm{\theta}}=\sqrt{\sigma^2_{\theta}/2}$. The parameters $\mu_{\theta}$ and $\sigma_{\theta}$ are the mean and standard deviation of the elevation angle, which are obtained based on the experimental measurements. For static users, they are reported as $\mu_{\theta}=41^\circ$ and $\sigma_{\theta}=7.68^\circ$. 
	Proof of \eqref{delta_val} is provided in Appendix~\ref{App-CDFCosPsi}. Furthermore, for the detailed proof of \eqref{LOSPDF}, we refer to Eq. (56) and (57) of \cite{MDSArxiv2018Orientation}. 
	The mean and scale factor of channel gain, $\mu_{\rm{H}}$ and $b_{\rm{H}}$ respectively, are:
	\begin{align}
	\label{LOSParameters}
	\mu_{\rm{H}} &=\frac{H_0}{d^{m+2}}\left(  \lambda_1 \sin{\mu_{\theta}} + \lambda_2 \cos{\mu_{\theta}}\right) , \\
	b_{\rm{H}} &=\frac{H_0}{d^{m+2}}b_{\rm{\theta}} |\lambda_1 \cos{\mu_{\theta}} - \lambda_2 \sin{\mu_{\theta}} | ,
	\end{align}
	where $H_0$ is given below \eqref{LOSSimple}. The factors, $\lambda_1$ and $\lambda_2$, are given in \eqref{Lambdas}. The support range of $f_{\rm{H}}(\hbar)$ is $h_{\rm{min}}\leq \hbar\leq h_{\rm{max}}$ where $h_{\rm{min}}$ and $h_{\rm{max}}$ are given as:
	\begin{equation}
	\label{hmin}
	h_{\rm{min}}=\begin{cases} \frac{H_0}{d^{m+2}}\cos\Psi_{\rm{c}}, &  \cos\psi<\cos\Psi_{\rm{c}}\\
	\dfrac{H_0}{d^{m+2}}\min\{\lambda_1,\lambda_2\} , & {\rm{o.w}}
	\end{cases} \ ,
	\vspace{-0.3cm}
	\end{equation}
	\begin{equation}
	\label{hmax}
	\!\!\!\!\!\!\!\!\!\!\!\!\!\!\!\!\!\!\!\!\!\!\!\!\!\!\!h_{\rm{max}}=\begin{cases}
	\dfrac{H_0}{d^{m+2}}\lambda_2 , & {\rm{if}}\ \ \  \lambda_1<0\\
	\dfrac{H_0}{d^{m+2}}\sqrt{\lambda_1^2+\lambda_2^2} , & {\rm{if}}\ \ \  \lambda_1\geq0 
	\end{cases} \ .
	\end{equation}
	
	The cumulative distribution function (CDF) of LOS channel gain can be also obtained by calculating the integral of \eqref{LOSPDF}, which is given as:
	\begin{equation}
	\begin{aligned}		
	\label{CDFLOSExact}
	&F_{\rm{H}}(\hbar)=c_{\rm{H}}+\\
	&\begin{cases}\!\dfrac{\exp\left(\frac{\hbar-\mu_{\rm{H}}}{b_{\rm{H}}} \right)-\exp\left(\frac{h_{\rm{min}}-\mu_{\rm{H}}}{b_{\rm{H}}} \right) }{\left(\!2-\exp\left(-\frac{h_{\rm{max}}-\mu_{\rm{H}}}{b_{\rm{H}}} \right)\!  \right) }, &h_{\rm{min}}\leq\hbar\leq\mu_{\rm{H}}\\
	\!\dfrac{2-\exp\left(\frac{h_{\rm{min}}-\mu_{\rm{H}}}{b_{\rm{H}}} \right)-\exp\left(-\frac{\hbar-\mu_{\rm{H}}}{b_{\rm{H}}} \right) }{\left(\!2-\exp\left(-\frac{h_{\rm{max}}-\mu_{\rm{H}}}{b_{\rm{H}}} \right)\!  \right) }, &h_{\rm{min}}\leq\mu_{\rm{H}}\leq\hbar\\
	\!\dfrac{\exp\left(-\frac{h_{\rm{min}}-\mu_{\rm{H}}}{b_{\rm{H}}} \right)- \exp\left(-\frac{\hbar-\mu_{\rm{H}}}{b_{\rm{H}}} \right)}{\left(\!2-\exp\left(-\frac{h_{\rm{max}}-\mu_{\rm{H}}}{b_{\rm{H}}} \right)\!  \right) }, &\mu_{\rm{H}}\leq h_{\rm{min}}\leq\hbar
	\end{cases} \ .
	\end{aligned}		
	\end{equation}

	

	
	\begin{figure*}[t!]
		\centering
		\subfloat[$\Omega=45^\circ$\label{sub1:figPDF}]{%
			\includegraphics[width=0.95\columnwidth,height=65mm,trim={0.5cm 0 -0.5cm 0}]{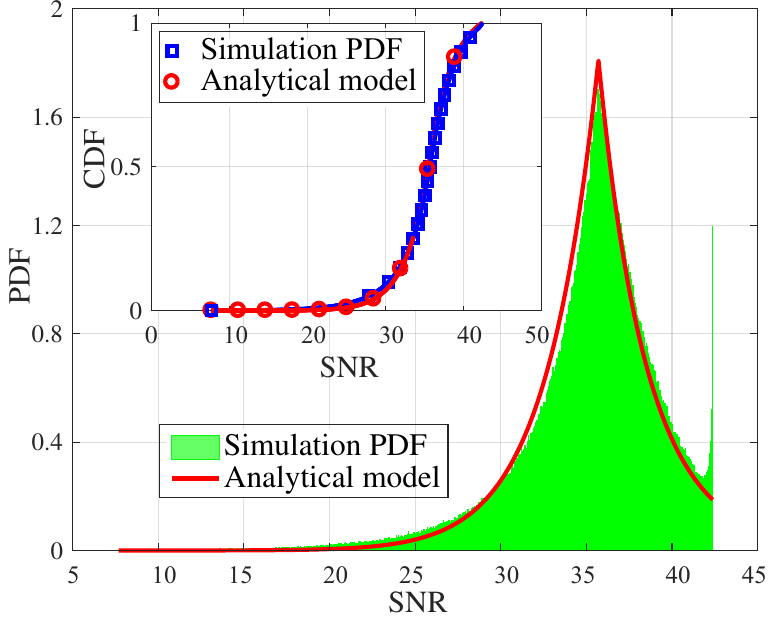}
		}\quad
		\subfloat[$\Omega=225^\circ$\label{sub2:figPDF}]{%
			\includegraphics[width=0.9\columnwidth,height=65mm]{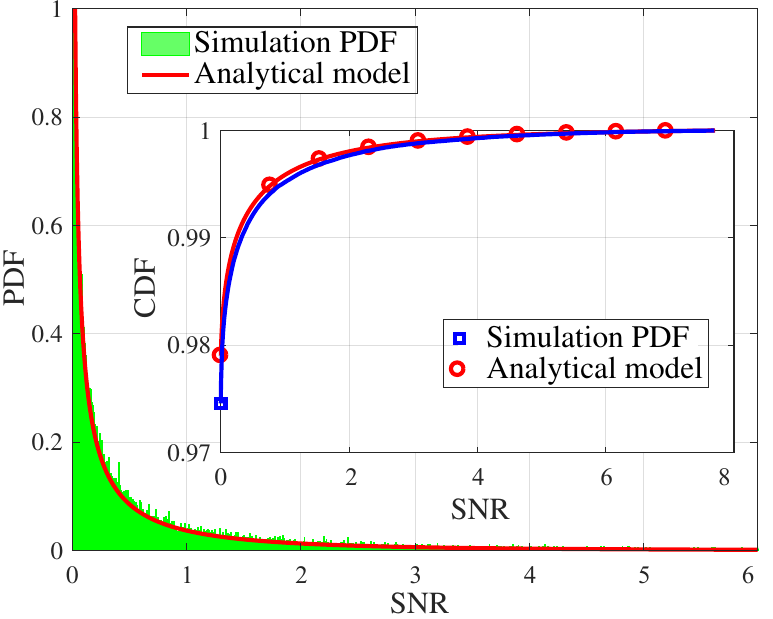}
		}\\
		\caption{Comparison between simulation and analytical results of PDF and CDF of received SNR for UE's location $L_1$ with $\Omega=45^{\circ}$ and $\Omega=225^{\circ}$. }
		\label{figPDF}
		\vspace{-0.2cm}
	\end{figure*}


	The relationship between channel gain and received SNR of DCO-OFDM is given in \eqref{SNR}. Using the fundamental theorem of determining the distribution of a random variable \cite{Papoulis}, the PDF of SNR can be obtained as follows:
	\begin{equation}
	\label{SNRPDF}
	\begin{aligned}		
	&f_{\mathcal{S}}(s)=\frac{f_H(\sqrt{s/\mathcal{S}_0})}{2\mathcal{S}_0\sqrt{s/\mathcal{S}_0}} \\
	&=\!\frac{\exp\left(-\frac{|\sqrt{s}-\sqrt{\mathcal{S}_0}\mu_{\rm{H}}|}{\sqrt{\mathcal{S}_0}b_{\rm{H}}} \right) }{2b_{\rm{H}}\sqrt{\mathcal{S}_0s}\left(2-\exp\left(-\frac{h_{\rm{max}}-\mu_{\rm{H}}}{b_{\rm{H}}} \right)  \right)}+c_{\rm{H}}\delta(s),
	\end{aligned}
	\end{equation}
	where $\mathcal{S}_0=\frac{R_{\rm{PD}}^2P_{\rm{opt}}^2}{(\mathcal{K}-2)\eta^2\sigma_k^2}$ and with the support range of $s\in(s_{\rm{min}},s_{\rm{max}})$, where $s_{\rm{min}}=\mathcal{S}_0h_{\rm{min}}^2$ and $s_{\rm{max}}=\mathcal{S}_0h_{\rm{max}}^2$, with $h_{\rm{min}}$ and $h_{\rm{max}}$ given in \eqref{hmin} and \eqref{hmax}, respectively.  
	
	By calculating the integral, $F_{\mathcal{S}}(s)=\int_{s_{\rm{min}}}^{s}f_{\mathcal{S}}(s){\rm{d}}s$, the CDF of SNR on $k$-th subcarrier can be obtained. The CDF of SNR can be also acquired by substituting $\hbar=\sqrt{\frac{s}{\mathcal{S}_0}}$ in \eqref{CDFLOSExact}, i.e., $F_{\mathcal{S}}(s)=F_{H}(\sqrt{\frac{s}{\mathcal{S}_0}})$.

	Fig.~\ref{figPDF} shows the PDF and CDF of the received SNR obtained from analytical results compared with the Monte-Carlo simulation results. The UE is located at position $L_1$, the transmitted optical power is $3.2$ W and UE's FOV is $90^{\circ}$. The results are provided for two directions: $\Omega=45^{\circ}$ and $\Omega=225^{\circ}$. Other simulation parameters are given in Table~\ref{TableSimulationParam}. As it can be seen, the analytical models for both PDF and CDF of the received SNR match the simulation results. The factor $c_{\rm{H}}$ for $\Omega=45^{\circ}$ is $0$. This factor for $\Omega=225^{\circ}$ is $0.975$ for simulation results and $0.979$ for analytical model. These results confirm the accuracy of the analytical model.
	
	\subsection{BER Performance}
	In this subsection, we aim to evaluate the effect of UE orientation on the BER performance of a LiFi-enabled device as one use case. BER is one of the common metrics to evaluate the point-to-point communication performance. Assuming the M-QAM DCO-OFDM modulation, the average BER per subcarrier of the communication link can be obtained as \cite{ghassemlooy2012optical}:
	\begin{equation}
	\label{BER}
	\bar{P}_{\rm{e}}=\int_{s_{\rm{min}}}^{s_{\rm{max}}} P_{\rm{e}}\left(s\right) f_{\mathcal{S}}(s)\ {\rm{d}}s , 
	\end{equation}
	where $P_{\rm{e}}$ determines the BER of $M$-QAM DCO-OFDM in additive white Gaussian noise (AWGN) channels, which can be obtained approximately as \cite{DimitrovClippingTCOM}:
	\begin{equation}
	\label{BERSNR}
	P_{\rm{e}}(s)\approx\frac{4}{\log_2M}\left(1-\frac{1}{\sqrt{M}} \right) Q\left(\sqrt{\frac{3s}{M-1}} \right),
	\end{equation}
	where $Q(\cdot)$ is the Q-function. Substituting \eqref{SNRPDF} and \eqref{BERSNR} into \eqref{BER} and calculating the integral from $s_{\rm{min}}$ to $s_{\rm{max}}$, we get the average BER of the $M$-QAM DCO-OFDM in AWGN channels with randomly-orientated UEs. 
	After calculating the integral and some simplifications, the approximated average BER is given as:
	\begin{equation}
	\label{BERApprox}
	\bar{P}_{\rm{e}}\approx\begin{cases} -\Delta_0  +\frac{1}{2}c_{\rm{H}}c_{\rm{M}}  , & \mu_{\rm{H}}\leq h_{\rm{min}}\\
	P_{\rm{e}}(\mathcal{S}_0\mu_{\rm{H}}^2)+\frac{1}{2}c_{\rm{H}}c_{\rm{M}} , & h_{\rm{min}}<\mu_{\rm{H}}\leq h_{\rm{max}}
	\end{cases} \ .
	\end{equation}
	where 
	\begin{equation}
	\begin{aligned}
	&\Delta_0=\frac{\frac{2}{\log_2M}\left(1-\frac{1}{\sqrt{M}} \right)\exp\left(\frac{\mu_{\rm{H}}-h_{\rm{min}}}{b_{\rm{H}}} \right)}{\left(2-\exp\left(-\frac{h_{\rm{max}}-\mu_{\rm{H}}}{b_{\rm{H}}} \right)  \right)}, \\
	&c_{\rm{M}}=\frac{4}{\log_2M}\left(1-\frac{1}{\sqrt{M}} \right). 
	\end{aligned}
	\end{equation}
	The proof is provided in Appendix~\ref{App-BER}. 
	
	\begin{figure}[t!]
		\begin{center}
			\resizebox{.95\linewidth}{!}{\includegraphics{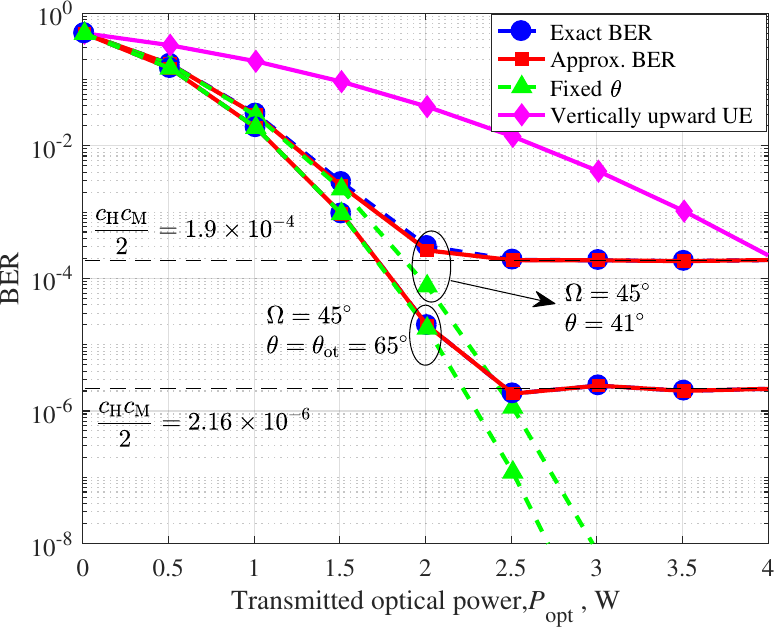}}
			\caption{BER performance of point-to-point communications for a UE located at $L_1$. Three scenarios are considered: i) vertically upward UE, ii) UE with the fixed polar angle without random orientation, and iii) real scenario with a random orientation (Laplace distribution) for polar angle.}
			\vspace{-0.4cm}
			\label{figBER}
		\end{center}
	\end{figure}

	Note that if the UE is tilted optimally towards the AP, the BER is minimum. For any arbitrary location and direction of UE, the optimum tilt (OT) angle is defined as the angle that provides maximum channel gain \cite{ImpactTilted,ICCTilting,jeong2013tilted,MDSWCNC19}. This angle is $\theta_{\rm{ot}}=\tan^{-1}\!\left(\!\frac{\lambda_1}{\lambda_2}\! \right)$ and the average BER for this tilt angle is $\bar{P}_{\rm{e}}\approx P_{\rm{e}}(\mathcal{S}_0\mu_{\rm{H}}^2)$ (since $c_{\rm{H}}=0$). 
	
	Fig.~\ref{figBER} illustrates the BER performance of $4$-QAM DCO-OFDM for three scenarios: i) a vertically upward UE, ii) a UE with a fixed polar angle and without random orientation iii) a realistic scenario in which the polar angle follows a Laplace distribution that considers the random orientation, i.e., $\theta\sim\mathcal{L}(\mu_{\theta},b_{\theta})$. Here, we assume $\mu_{\theta}=41^\circ$ and $b_{\theta}=5.43^\circ$ as reported in\cite{MDSArxiv2018Orientation} based on the experimental measurements. Other simulation parameters are given in Table~\ref{TableSimulationParam}. The results are provided for the UE's location of $L_1=(3,3)$ and with $\Psi_{\rm c}=60^{\circ}$. For this location, $\theta_{\rm{ot}}\approx65^{\circ}$. Some interesting observations can be seen from the results shown in this figure. 
	As can be seen, for $\Omega=45^\circ$, the vertically upward UE falls behind the other two scenarios. Because for $\theta>0$, the UE will be tilted towards the AP (see the results shown in Fig.~\ref{figCEOmega}). Also, the gap between the exact and approximate BER is small which confirms the accuracy of the BER approximation. 
	One interesting observation is that after $P_{\rm{opt}}>2$ W and $P_{\rm{opt}}>2.5$ W, the BER does not decrease and is saturated for $\theta=41^{\circ}$ and $\theta=\theta_{\rm ot}$, respectively. This is due to the constant term in \eqref{BERApprox}, i.e., $\frac{1}{2}c_{\rm{H}}c_{\rm{M}}$, will be dominant compared to the power-dependent term, i.e., $P_{\rm{e}}(\mathcal{S}_0\mu_{\rm{H}}^2)$. In other words, due to the random orientation, there are cases that LOS link is out of the UE's FOV and data is lost. These results highlight the significance of considering the random orientation in the performance assessment. 
	The BER performance of second and third scenarios can still be better if $\theta=\theta_{\rm{ot}}\approx65^{\circ}$. 
	For $\theta=\theta_{\rm{ot}}$ the maximum LOS channel gain is achieved and under this condition the BER is minimum. This fact underlines that the device orientation is not always destructive. Furthermore, with $\theta=\theta_{\rm{ot}}$ the UE's random orientation has the minimum effect on the BER. 
	We note that for a given location and $\Omega$, the $\bar{P}_{\rm{e}}$ given in \eqref{BERApprox} is always bounded to the BER of $P_{\rm{e}}(s)$ obtained for $\theta=\theta_{\rm{ot}}$ as it provides the maximum LOS channel gain. The BER results of $\theta=\theta_{\rm{ot}}$ are just provided for a comparison purpose however, the users tends to keep their smartphone with $\theta=41^{\circ}$ (when doing sitting activities) according to the experimental measurements \cite{MDSArxiv2018Orientation}. 

	\subsection{UE's Random Motion}
	In this subsection, we will include the effect of UE's random motion even though the user is static in addition to the random orientation on the BER performance. Note that here, the random UE's motion encompass small movements in $x$, $y$ and $z$ directions, which are modeled as Gaussian distributions. Hence, the UE's location at each realization is given as:
	\begin{equation}
	(x_{\rm u},y_{\rm u},z_{\rm u})=(x_{\rm 0,u},y_{\rm 0,u},z_{\rm 0,u})+(\Delta x,\Delta y,\Delta z),
	\end{equation}
	where $\Delta x\sim\mathcal{N}(x_{\rm 0,u},\sigma_{\rm x}^2)$, $\Delta y\sim\mathcal{N}(y_{\rm 0,u},\sigma_{\rm y}^2)$ and $\Delta z\sim\mathcal{N}(z_{\rm 0,u},\sigma_{\rm z}^2)$. The location $(x_{\rm 0,u},y_{\rm 0,u},z_{\rm 0,u})$ denotes the mean point that the UE fluctuates around. It is noted that typically the variation of the UE's height (along $z$ axis) is less than the variation along $x$ and $y$ axes. 
	
	\begin{figure}[t!]
		\begin{center}
			\resizebox{.95\linewidth}{!}{\includegraphics{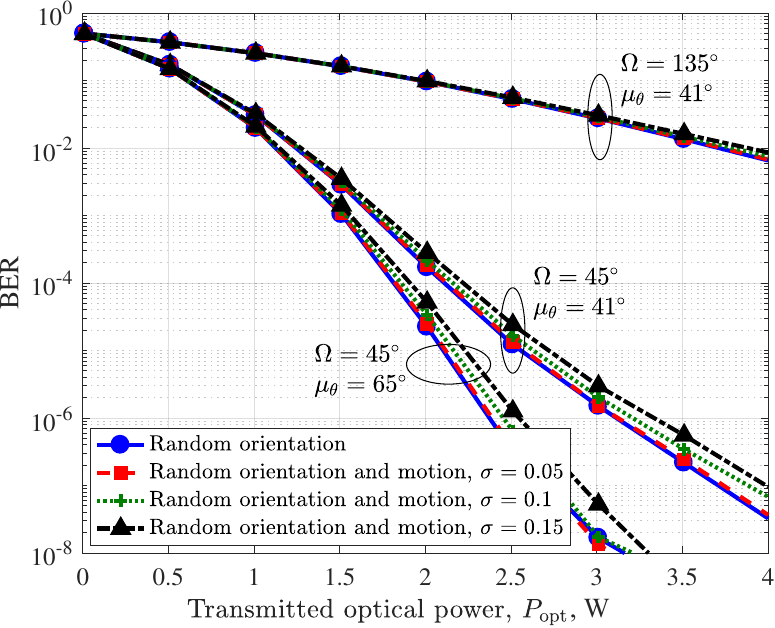}}
			\caption{The effect of random orientation with/without random motion on BER performance of a UE located at the arbitrary position of $L_1$.}
			\vspace{-0.45cm}
			\label{BERxyz}
		\end{center}
	\end{figure}
	
	Fig. \ref{BERxyz} shows the effect of random motion along with random orientation on the BER performance. In these simulations, we assume that $\sigma_{\rm x}=\sigma_{\rm y}=5\sigma_{\rm z}=\sigma$. The results are presented for three values of $\sigma$, which are $0.05$ m, $0.1$ m and $0.15$ m. Note that for $\sigma=0.15$ m, the deviation of the UE's location from the mean point, $(x_{\rm 0,u},y_{\rm 0,u},z_{\rm 0,u})$, can be in the range of $-3\sigma=-45$ cm to $3\sigma=45$ cm. This corresponds to high UE's motion which is very low probable for normal human activities. Here, the modulation order is considered to be $M=4$. The simulations are carried out for a UE located at $L_1=(3,3)$ with different $\Omega$ and $\mu_{\theta}$. The UE's FOV is assumed to be $90^{\circ}$ for these simulations. 
	As it can be seen, with $\sigma\in\{0.05,0.1\}$, the gap between the results when random motion is included, is indeed negligible. For the case of $\Omega=135^{\circ}$ and $\mu_{\theta}=41^{\circ}$ and with $\sigma=0.15$ m, the gap is still small. 
	For $\Omega=45^{\circ}$ and $\mu_{\theta}=41^{\circ}$ (or $\mu_{\theta}=65^{\circ}$) with $\sigma=0.15$ m, the gap grows in high transmitted power.

	\begin{figure}[t!]
		\begin{center}
			\resizebox{.65\linewidth}{!}{\includegraphics{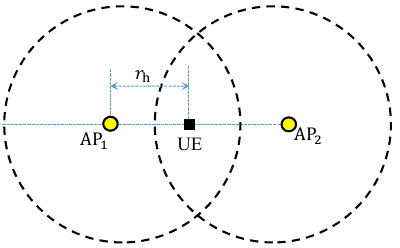}}
			\caption{ Geometry of two APs with interference consideration. APs are located at $(-2,0)$ and $(2,0)$ on the ceiling.  }
			\vspace{-0.45cm}
			\label{SINR}
		\end{center}
	\end{figure}
	
	\begin{figure}[t!]
		\begin{center}
			\resizebox{.95\linewidth}{!}{\includegraphics{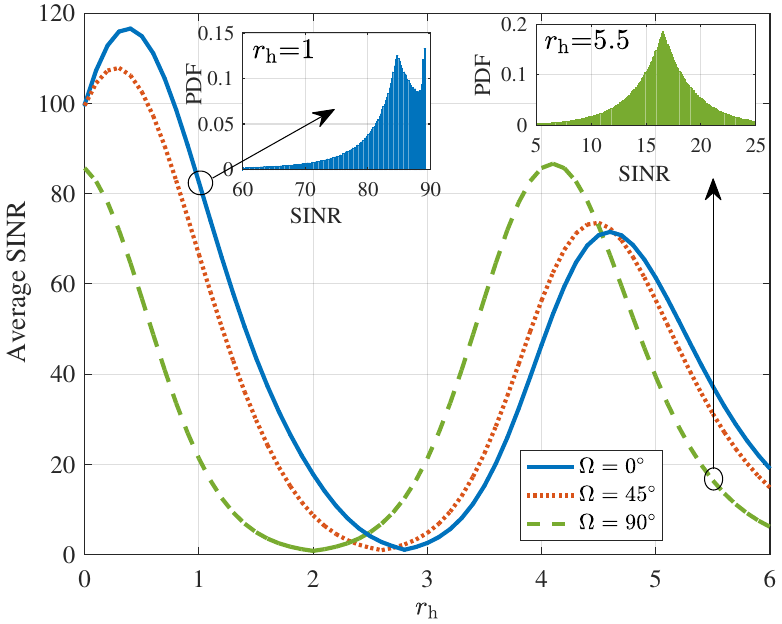}}
			\caption{ Average SINR versus the horizontal distance of the UE and first AP, $r_{\rm h}$, (see the geometry shown in Fig.~\ref{SINR}).  }
			\vspace{-0.45cm}
			\label{SINR_res}
		\end{center}
	\end{figure}
	
	\subsection{Multiple APs Scenario}
	To investigate the effect of multiple APs on the error performance of a randomly-orientated UE, we consider two APs located at $(-2,0)$ and $(2,0)$ as shown in Fig.~\ref{SINR}. The signal-to-interference- plus-noise ratio (SINR) can be obtained as:
	\begin{equation}
	\Upsilon=\frac{R_{\rm{PD}}^2H_{\rm d}^2P_{\rm{opt}}^2}{(\mathcal{K}-2)\eta^2(\sigma_k^2+I)},
	\end{equation}
	where $H_{\rm d}$ is the LOS channel gain between the desired AP and the PD; $I$ is the interfering power from other APs on the $k$th subcarrier. Other parameters are defined below (13). Here, with the consideration of two APs, the interference from the other AP on the $k$th subcarrier is $I=R_{\rm{PD}}^2H_{\rm in}^2P_{\rm{opt}}^2/((\mathcal{K}-2)\eta^2)$, where $H_{\rm in}$ is the channel gain between the interfering AP and the UE. Note that the desired AP is selected based on the received signal intensity metric. 
	Fig.~\ref{SINR_res} shows the average SINR versus different horizontal distances between the UE and first AP (as depicted in Fig.~\ref{SINR}). The average is taken over different random orientations following a Laplace distribution based on the experimental measurements, i.e., $\theta\sim\mathcal{L}(41^\circ,5.43^\circ)$. Note that mobility is not considered in these results and at each location the user is assumed to be sitting. The simulation parameters are given in Table~\ref{TableSimulationParam} and the UE's FOV is assumed to be $90^{\circ}$. The transmitted optical power per AP is supposed to be $1$ W as multiple APs require lower transmit power to cover the room in comparison to the single AP case. The PDF of SINR for $r_{\rm h}=1$ with $\Omega=0$ and $r_{\rm h}=4.5$ with $\Omega=90^{\circ}$ are presented. For the former the average SINR is about $82$ while for the latter, it is about $16$. 
	Note also that the PDF of SINR shows similar Laplacian distributions as in the SNR case.


	

	\section{Conclusions and Future Works}
	\label{ConclFuture}
	We analyzed the device orientation and assessed its importance on system performance. The PDF of SNR for randomly-orientated device is derived, and based on the derived PDF, the BER performance of DCO-OFDM in AWGN channel with randomly-orientated UEs is evaluated. An approximation for the average BER of randomly-oriented UEs is calculated that closely matches the exact one. The role of CE angle that guarantees having LOS link in the UE's FOV is investigated. Furthermore, the significant impact of being optimally tilted towards the AP on the BER performance is shown. We also studied the effect of the UE's random motion on the BER performance. 
	We note that even though we considered DCO-OFDM, the methodology can be readily extended to other modulation schemes, which can be the focus of future studies. Furthermore, other performance metrics such as throughput and user's quality of service can also be assessed. Also, the device orientation impact can be evaluated in a cellular network with consideration of non-line-of-sight links.
	
	\section*{Acknowledgment}
	M.D. Soltani acknowledges the School of Engineering for providing financial support. Harald Haas and Majid Safari gratefully acknowledge financial support from EPSRC under grant EP/L020009/1 (TOUCAN). 
	
	\appendix
	\subsection{Proof of \eqref{CEComplete}}
	\label{App-CE}
	Recalling that $\cos\psi=-{\bf{d}}\cdot\mathbf{n}'_{\rm{u}}/d$, replacing for ${\bf{d}}=[x_{\rm{u}}-x_{\rm{a}},y_{\rm{u}}-y_{\rm{a}},z_{\rm{u}}-z_{\rm{a}}]^{\rm{T}}$ and 
	$\mathbf{n}'_{\rm{u}}=[\sin\theta\cos\omega,\sin\theta\sin\omega,\cos\theta]^{\rm{T}}$ and also noting that $\omega=\Omega+\pi$, we have:
	\begin{equation}
	\begin{aligned}
	\label{Cospsitheta}
	&\resizebox{1\hsize}{!}{$\cos\psi=\dfrac{\!\!(x_{\rm{u}}-x_{\rm{a}})\!\sin\theta\cos\Omega+(y_{\rm{u}}-y_{\rm{a}})\!\sin\theta\sin\Omega-(z_{\rm{u}}-z_{\rm{a}})\!\cos\theta}{\sqrt{(x_{\rm{u}}-x_{\rm{a}})^2+(y_{\rm{u}}-y_{\rm{a}})^2+(z_{\rm{u}}-z_{\rm{a}})^2}}$}\\
	&\!\resizebox{0.025\hsize}{!}{$=$}\dfrac{\!\!\resizebox{0.97\hsize}{!}{$\sqrt{(x_{\rm{u}}-x_{\rm{a}})^2\!+(y_{\rm{u}}-y_{\rm{a}})^2}\sin\theta\cos\!\left(\! \Omega\!-\tan^{-1}\!\left(\!\dfrac{y_{\rm{u}}-y_{\rm{a}}}{x_{\rm{u}}-x_{\rm{a}}}\! \right)\!\! \right)\!\!-(z_{\rm{u}}\!-z_{\rm{a}})\!\cos\theta $} }{\sqrt{(x_{\rm{u}}-x_{\rm{a}})^2+(y_{\rm{u}}-y_{\rm{a}})^2+(z_{\rm{u}}-z_{\rm{a}})^2}}\\
	&=\frac{r}{d}\sin\theta\cos\!\left(\! \Omega\!-\tan^{-1}\!\left(\!\dfrac{y_{\rm{u}}-y_{\rm{a}}}{x_{\rm{u}}-x_{\rm{a}}}\! \right)\!\! \right)+\frac{h}{d}\cos\theta .
	\end{aligned}
	\end{equation}
	For a given location of UE and a fixed angle of $\Omega$, by using the simple triangular rules, $\cos\psi$ can be represented as:
	\begin{equation}
	\label{CospsiTheta}
	\resizebox{1\hsize}{!}{$\cos\psi=\!\lambda_1\sin\theta+\lambda_2\cos\theta=\!\sqrt{\lambda_1^2+\lambda_2^2}\cos\left(\!\theta-\tan^{-1}\left(\dfrac{\lambda_1}{\lambda_2} \right)\!\!\right),$}
	\end{equation}
	where $\lambda_1$ and $\lambda_2$ are given as:
	\begin{equation}
	\begin{aligned}
	&\lambda_1=\frac{r}{d}\cos\left(\! \Omega-\tan^{-1}\!\left(\frac{y_{\rm{u}}-y_{\rm{a}}}{x_{\rm{u}}-x_{\rm{a}}} \right)\! \right),\\
	&\lambda_2=\frac{h}{d}.
	\end{aligned}
	\end{equation}
	According to the definition of critical elevation angle, if $\theta=\theta_{\rm{ce}}$, then, $\cos\psi=\cos\Psi_{\rm{c}}$. Therefore, \eqref{CospsiTheta} results in:
	\begin{equation}
	\label{ThetaCE}
	\theta_{\rm{ce}}=\cos^{-1}\left(\frac{\cos\Psi_{\rm{c}}}{\sqrt{\lambda_1^2+\lambda_2^2}} \right)+\tan^{-1}\left(\frac{\lambda_1}{\lambda_2} \right).
	\end{equation}
	This completes the proof of the derivation of CE angle. 
	
	\subsection{Proof of Proposition}
	\label{App-Proposition}
	For a given location of UE and a fixed elevation angle, one other representation of $\cos\psi$ given in \eqref{Cospsitheta} would be as a function of $\Omega$:\vspace{-0.1cm}
	\begin{equation}
	\begin{aligned}
	\label{CospsiOmega}
	\cos\psi\!=\!
	\kappa_1\cos\!\left(\!\Omega\!-\tan^{-1}\!\left(\!\frac{y_{\rm{u}}-y_{\rm{a}}}{x_{\rm{u}}-x_{\rm{a}}}\! \right)\!\right)\!+\kappa_2\triangleq \Lambda(\Omega),
	\end{aligned}
	\end{equation}
	where the coefficients $\kappa_1$ and $\kappa_2$ are given as:
	\begin{equation}
	\begin{aligned}
	\kappa_1=\frac{r}{d}\sin\theta, \ \ \ \ 
	\kappa_2=\frac{h}{d}\cos\theta.
	\end{aligned}
	\end{equation}
	Note that since $\theta\in[0,90]$, we have $\kappa_1\geq 0$ and $\kappa_2\geq 0$. 
	As mentioned for $\theta=\theta_{\rm{ce}}$, we have $\cos\psi=\cos\Psi_{\rm{c}}$. Then, solving $\Lambda(\Omega)-\cos\Psi_{\rm{c}}=0$ for $\Omega$, the roots are  $\Omega_{\rm{r1}}=\min\{\Omega_1,\Omega_2\}$ and $\Omega_{\rm{r2}}=\max\{\Omega_1,\Omega_2\}$, where $\Omega_1$ and $\Omega_2$  are given as follow:
	\begin{equation}
	\begin{aligned}
	\label{Roots}
	&\Omega_1=\cos^{-1}\!\left(\!\frac{\cos\Psi_{\rm{c}}-\kappa_2}{\kappa_1}\!\right)+\!\tan^{-1}\!\!\left(\!\frac{y_{\rm{u}}-y_{\rm{a}}}{x_{\rm{u}}-x_{\rm{a}}}\! \right),\\
	&\Omega_2=-\cos^{-1}\!\left(\!\frac{\cos\Psi_{\rm{c}}-\kappa_2}{\kappa_1}\!\right)+\!\tan^{-1}\!\!\left(\!\frac{y_{\rm{u}}-y_{\rm{a}}}{x_{\rm{u}}-x_{\rm{a}}}\! \right).
	\end{aligned}
	\end{equation}
	For the special case of $\Psi_{\rm{c}}=90^\circ$, \eqref{Roots} is simplified as:
	\begin{equation}
	\begin{aligned}
	&\Omega_1\!=\!\cos^{-1}\left(\frac{-h\cot\theta}{r}\right)+\tan^{-1}\!\left(\!\frac{y_{\rm{u}}-y_{\rm{a}}}{x_{\rm{u}}-x_{\rm{a}}}\! \right),\\
	&\Omega_2\!=\!-\cos^{-1}\left(\frac{-h\cot\theta}{r}\right)+\tan^{-1}\!\left(\!\frac{y_{\rm{u}}-y_{\rm{a}}}{x_{\rm{u}}-x_{\rm{a}}}\! \right).
	\end{aligned}
	\end{equation}
	Using the sinuous function properties if $\Lambda(\Omega)\leq0$ for $\Omega\in[\Omega_{\rm{r1}},\Omega_{\rm{r2}}]$, then the derivative of $\Lambda(\Omega)$ at $\Omega=\Omega_{\rm{r1}}$ is negative, i.e.,
	$\frac{\partial \Lambda(\Omega)}{\partial \Omega}|_{\Omega=\Omega_{\rm{r1}}}<0$. 
	For simplicity of notation, let's denote $\Lambda'(\Omega)=\frac{\partial \Lambda(\Omega)}{\partial \Omega}$. Using \eqref{CospsiOmega},  we have $\Lambda'(\Omega)=-\kappa_1\sin\left(\Omega-\tan^{-1}\left(\frac{y_{\rm{u}}-y_{\rm{a}}}{x_{\rm{u}}-x_{\rm{a}}} \right)\right)+\kappa_2$. 
	Therefore, the range of $\mathcal{R}_{\Omega}$ that guarantees $\Lambda(\Omega)>0$ would be $[0,\Omega_{\rm{r1}})\bigcup(\Omega_{\rm{r2}},2\pi]$. Similarly, if $\Lambda(\Omega)\geq0$ for $\Omega\in(\Omega_{\rm{r1}},\Omega_{\rm{r2}})$, then the derivative of $\Lambda(\Omega)$ at $\Omega=\Omega_{\rm{r1}}$ is positive, i.e.,
	$\frac{\partial \Lambda(\Omega)}{\partial \Omega}|_{\Omega=\Omega_{\rm{r1}}}>0$. Consequently, in this case the range of $\mathcal{R}_{\Omega}$ that ensures $\Lambda(\Omega)>0$ would be $[\Omega_{\rm{r1}},\Omega_{\rm{r2}}]$. This completes the proof of Proposition.

	\subsection{Proof of \eqref{delta_val}}
	\label{App-CDFCosPsi}
	Using \eqref{CospsiTheta}, the CDF of $\cos\psi$ can be obtained as:
	\begin{equation}
	\label{CDFCospsi}
	\begin{aligned}
	&F_{\rm{\cos\psi}}(\tau)=\Pr\{\cos\psi\leq\tau\}\\
	&=\Pr\left\lbrace \sqrt{\lambda_1^2+\lambda_2^2}\cos\left(\!\theta-\tan^{-1}\left(\dfrac{\lambda_1}{\lambda_2} \right)\!\!\right)\leq\tau\right\rbrace \\
	&=1-F_{\theta}\left(\cos^{-1}\left(\frac{\tau}{\sqrt{\lambda_1^2+\lambda_2^2}} \right)+\tan^{-1}\left(\frac{\lambda_1}{\lambda_2} \right)   \right).
	\end{aligned}
	\end{equation}
	where $F_{\theta}(\theta)$ is the CDF of the elevation angle, $\theta$. Under the assumption of Laplacian model for the elevation angle, $F_{\theta}(\theta)$ is given as \cite{MDSArxiv2018Orientation}:
	\begin{equation}
	F_{\theta}(\theta)=\begin{cases}
	\frac{1}{2\left(G(\frac{\pi}{2})-G(0) \right)}\exp\left(\frac{\theta-\mu_{\theta}}{b_{\rm{\theta}}} \right),   \ \ \ \  &\theta<\mu_{\theta} \\
	1\!-\frac{1}{2\left(G(\frac{\pi}{2})-G(0) \right)}\exp\!\left(-\frac{\theta-\mu_{\theta}}{b_{\rm{\theta}}} \right),   &\theta\geq\mu_{\theta}
	\end{cases}.
	\end{equation}
	where $G(0)\!=\!\frac{1}{2}\exp\left(\frac{-\mu_{\theta}}{b_{\rm{\theta}}} \right)$ and $G(\frac{\pi}{2})=1-\frac{1}{2}\exp\left(-\frac{\frac{\pi}{2}-\mu_{\theta}}{b_{\rm{\theta}}} \right)$. Note that with reported values for $\mu_{\theta}$ and $b_{\rm{\theta}}$ from \cite{MDSArxiv2018Orientation}, we have $\left(G(\frac{\pi}{2})-G(0) \right)\approx1$. Therefore, 
	\begin{equation}
	F_{\theta}(\theta)\approx\begin{cases}
	\frac{1}{2}\exp\left(\frac{\theta-\mu_{\theta}}{b_{\rm{\theta}}} \right),   \ \ \ \ \  &\theta<\mu_{\theta} \\
	1-\frac{1}{2}\exp\left(-\frac{\theta-\mu_{\theta}}{b_{\rm{\theta}}} \right),   \ \ \  &\theta\geq\mu_{\theta}
	\end{cases}.
	\end{equation}
	Finally, by recalling the definition of the CE angle given in \eqref{CEComplete}, $F_{\rm{\cos\psi}}(\cos\Psi_{\rm{c}})$ can be approximately obtained as:
	\begin{equation}
	\label{delta_valproof}
	F_{\rm{\cos\psi}}(\cos\Psi_{\rm{c}})\!\approx\! \begin{cases}
	1-\frac{1}{2}\exp\left( \frac{\theta_{\rm{ce}}-\mu_{\theta}}{b_{\theta}}\right),   &\theta_{\rm{ce}}<\mu_{\theta}\\
	\frac{1}{2}\exp\left( -\frac{\theta_{\rm{ce}}-\mu_{\theta}}{b_{\theta}}\right),  &\theta_{\rm{ce}}\geq\mu_{\theta} 
	\end{cases} \ .
	\end{equation}
	This completes the proof of \eqref{delta_val}. 
	
	\subsection{Proof of \eqref{BERApprox}}
	\label{App-BER}
	Substituting \eqref{SNRPDF} and \eqref{BERSNR} into \eqref{BER}, we have:
	\begin{equation}
	\label{BER1}
	\begin{aligned}
	\bar{P}_{\rm{e}}&\!=c_0\!\!\int_{s_{\rm{min}}}^{s_{\rm{max}}}\!\!Q\left(\! \sqrt{\frac{3s}{M-1}}\right)\frac{1}{\sqrt{s}}\exp\!\left(-\frac{\!|\sqrt{s}-\sqrt{\mathcal{S}_0}\mu_{\rm{H}}|}{\sqrt{\mathcal{S}_0}b_{\rm{H}}} \right) {\rm{d}}s\\
	&+c_{\rm{H}}c_{\rm{M}} \int_{s_{\rm{min}}}^{s_{\rm{max}}}Q\left(\sqrt{\frac{3s}{M-1}}\right)\delta(s){\rm{d}}s
	\end{aligned}
	\end{equation}
	with $c_0$ and $c_{\rm{M}}$ given as:
	\begin{equation}
	\begin{aligned}
	&c_0=\frac{c_{\rm{M}}}{2b_{\rm{H}}\sqrt{\mathcal{S}_0}\left(2-\exp\left(-\frac{h_{\rm{max}}-\mu_{\rm{H}}}{b_{\rm{H}}} \right)  \right)},\\
	&c_{\rm{M}}=\frac{4}{\log_2M}\left(1-\frac{1}{\sqrt{M}} \right). 
	\end{aligned}
	\end{equation}
	Note that if $s_{\rm{min}}=0$, the second integral in \eqref{BER1} is $c_{\rm{H}}c_{\rm{M}}Q(0)=\frac{c_{\rm{H}}c_{\rm{M}}}{2}$, and referring to the definition of $c_{\rm{H}}$, it is zero for $s_{\rm{min}}>0$. Thus, the second integral can be expressed as $\frac{c_{\rm{H}}c_{\rm{M}}}{2}$ and we need to simplify the first integral. 
	For simplicity of notation, let define $c_1=\sqrt{\frac{3}{M-1}}$, $c_2=\sqrt{\mathcal{S}_0}\mu_{\rm{H}}$ and $c_3=\sqrt{\mathcal{S}_0}b_{\rm{H}}$. Furthermore, let $x=\sqrt{s}$, thus, the first integral in \eqref{BER1} can be rewritten as \eqref{BER2} given at the top of the next page.   
	\begin{figure*}[!t]
		\begin{equation}
		\label{BER2}
		\begin{aligned}
		&\int_{\sqrt{s_{\rm{min}}}}^{\sqrt{s_{\rm{max}}}}Q(c_1x)e^{-\frac{|x-c_2|}{c_3}} {\rm{d}}x=\begin{cases}
		\int_{\sqrt{s_{\rm{min}}}}^{\sqrt{s_{\rm{max}}}}Q(c_1x)e^{-\frac{x-c_2}{c_3}} {\rm{d}}x, & c_2\leq\sqrt{s_{\rm{min}}} \vspace{0.3cm}\\
		\int_{\sqrt{s_{\rm{min}}}}^{c_2}Q(c_1x)e^{\frac{x-c_2}{c_3}}{\rm{d}}x +\int_{c_2}^{\sqrt{s_{\rm{max}}}}Q(c_1x)e^{-\frac{x-c_2}{c_3}} {\rm{d}}x, 
		& \sqrt{s_{\rm{min}}}< c_2\leq\sqrt{s_{\rm{max}}}
		\end{cases},\\& 
		\end{aligned}
		\end{equation}
		\vspace{-0.4cm}
		\hrule
		\vspace{-0.2cm}
	\end{figure*}
	The right side of \eqref{BER2} is based on the behavior of PDF of SNR. It can be either single exponential (if $c_2\geq\sqrt{s_{\rm{min}}}$) or double exponential (if $\sqrt{s_{\rm{min}}}< c_2\leq\sqrt{s_{\rm{max}}}$), for example, see results shown in Fig. \ref{figPDF}.
	Noting that 
	\begin{equation}
	\begin{aligned}
	&\int Q(c_1x)e^{\frac{x}{c_3}}{\rm{d}}x=\\
	&c_3e^{\frac{x}{c_3}}Q(c_1x)+\frac{c_3}{2}e^{\frac{1}{4c_1^2c_3^2}}\left(1-2Q\left(c_1x-\frac{1}{2c_1c_3}\right) \right) ,
	\end{aligned}
	\end{equation}
	also for given values of $c_1$, $c_2$ and $c_3$, we have $Q(c_1c_2)\approx Q(c_1\sqrt{s_{\rm{max}}})$ and also since $\mu_{\rm H}>>b_{\rm H}$, then, $e^{-\frac{c_2}{c_3}}\approx0$. Hence, $\bar{P}_{\rm{e}}$ can be approximated by \eqref{BERApprox1} presented at the top of the next page.
	\begin{figure*}[!t]
		\begin{equation}
		\label{BERApprox1}
		\bar{P}_{\rm{e}}\approx\begin{cases}-c_0c_3 +\frac{c_{\rm{H}}c_{\rm{M}}}{2},   &c_2\leq \sqrt{s_{\rm{min}}}\\
		2c_0c_3Q(c_1c_2)\left(2-e^{\frac{c_2-\sqrt{s_{\rm{max}}}}{c_3}} \right)+\frac{c_{\rm{H}}c_{\rm{M}}}{2},  &\sqrt{s_{\rm{min}}}< c_2\leq\sqrt{s_{\rm{max}}}
		\end{cases} \ .
		\end{equation}
		\hrule
	\end{figure*}
	By substituting for the values of $c_0$, $c_1$, $c_2$, $c_3$ and noting that $\sqrt{s_{\rm{min}}}=\sqrt{\mathcal{S}_0}h_{\rm{min}}$ and $\sqrt{s_{\rm{max}}}=\sqrt{\mathcal{S}_0}h_{\rm{max}}$ \eqref{BERApprox1} can be rewritten as:
	\begin{equation}
	\label{BERApprox2}
	\bar{P}_{\rm{e}}\!\approx\!\begin{cases}\!-\frac{\frac{2}{\log_2M}\left(\!1-\frac{1}{\sqrt{M}}\! \right)e^{\frac{\mu_{\rm{H}}-h_{\rm{min}}}{b_{\rm{H}}} }}{\left(2-\exp\left(-\frac{h_{\rm{max}}-\mu_{\rm{H}}}{b_{\rm{H}}} \right)  \right)}+\frac{c_{\rm{H}}c_{\rm{M}}}{2}  , \ \  \mu_{\rm{H}}\leq h_{\rm{min}}\vspace{0.3cm}\\
	\frac{4\left(1-\frac{1}{\sqrt{M}} \right)}{\log_2M} Q\left(\sqrt{\frac{3\mathcal{S}_0\mu_{\rm{H}}^2}{M-1}} \right)+\frac{c_{\rm{H}}c_{\rm{M}}}{2}  , \   h_{\rm{min}}<\mu_{\rm{H}}\leq h_{\rm{max}}
	\end{cases} \ .
	\end{equation}
	This completes the proof of \eqref{BERApprox}.

	\footnotesize
	\bibliographystyle{IEEEtran}
	\bibliography{IEEEabrv.bib,Ref.bib}

\end{document}